\documentclass[]{spie}  
\usepackage[]{graphicx}
\usepackage[]{times}

\title{A multi-object, multi-field spectrometer and imager for a European ELT} 

\author{
Chris Evans\supit{a}, Colin Cunningham\supit{a}, Eli
Atad-Ettedgui\supit{a}, Jeremy Allington-Smith\supit{b}, Francois
Ass\'{e}mat\supit{b}, Gavin Dalton\supit{c,d}, Peter Hastings\supit{a},
Timothy Hawarden\supit{a}, Isobel Hook\supit{c}, Rob Ivison\supit{a},
Simon Morris\supit{b}, Suzanne Ramsay Howat\supit{a}, Mel Strachan\supit{a}
and Stephen Todd\supit{a}
\skiplinehalf
\supit{a}UK Astronomy Technology Centre, Royal Observatory, Blackford Hill, Edinburgh, EH9 3HJ, UK; \\
\supit{b}Dept. of Physics, University of Durham, South Road, DH1 3LE, UK; \\
\supit{c}Dept. of Astrophysics, Denys Wilkinson Building, Keble Road, Oxford, OX1 3RH, UK; \\
\supit{d}Space Science \& Technology Dept., CCLRC Rutherford Appleton Lab., Chilton, OX11~0QX, UK \\
}

\begin{document} 
\maketitle

\begin{abstract}
One of the highlights of the European ELT Science Case book is the
study of resolved stellar populations, potentially out to the Virgo
Cluster of galaxies.  A European ELT would enable such studies in a
wide range of unexplored distant environments, in terms of both
galaxy morphology and metallicity.  As part of a small study, a
revised science case has been used to shape the conceptual design of a
multi-object, multi-field spectrometer and imager (MOMSI).  Here we
present an overview of some key science drivers, and how to achieve
these with elements such as multiplex, AO-correction,
pick-off technology and spectral resolution.
\end{abstract}



\section{INTRODUCTION}\label{intro}  

The possibility of studying resolved stellar populations in the Virgo
cluster of galaxies (at $\sim$16~Mpc) was one of the foremost sections
of the OPTICON Science Case for a European Extremely Large Telescope
\cite{imh} (E-ELT).  A very large primary aperture combined
with improved spatial resolution over that from 10-m class facilities
offers immense potential for studies of stellar populations in
galaxies, from the edge of the Local Group right out to Virgo.

As part of the FP6 Instrument Small Studies for an E-ELT we have
revisited the science case and developed an initial design concept for
a multi-object, multi-field spectrometer and imager (MOMSI).  As the
plans for an E-ELT have evolved, the relative strengths of different
science cases have also evolved.  Here we present potential studies of
resolved stellar populations that would benefit from an E-ELT,
especially when combined with significant wavefront correction from an
adaptive optics (AO) system.  Due to the technical
drive toward the near-IR (in which the AO correction is more
effective) a MOMSI-type instrument would obviously also benefit studies of
galaxy evolution and other programmes at high redshift; for
completeness we highlight a couple of areas to which MOMSI could
contribute.  Although less demanding in terms of the required
AO-correction, we note that one of the most compelling science cases for a
multi-object near-IR spectrograph on an ELT will most likely concern
the follow-up of faint galaxies discovered by the {\it James Webb
Space Telescope (JWST)}.

\section{SCIENCE CASE}

Before discussing specific cases it is worth identifying a
more general requirement from the desire to study resolved stellar
populations with an E-ELT.  The correction from adaptive optics (AO)
becomes more effective at longer wavelengths; the reality of
significant AO-correction in the optical domain may yet be some time
ahead.  Many conceivable projects to study resolved stellar
populations are still dependent on optical diagnostics at some level
-- some of this is perhaps historical, but it is also true that there
are generally fewer diagnostics in the near-IR, both photometrically
and spectroscopically.  In this context, access to the $I$ and $Z$
passbands would be of significant benefit.  Photometry in these bands
adds significant leverage on age and metallicity determinations of
e.g. the main-sequence turn-off in old populations.  Moreover,
spectroscopy in the $I$-band region gives access to the Calcium II
triplet (CaT), with rest wavelengths of 850, 854, and 866~nm.  In
contemporary studies of stellar kinematics the CaT is the fundamental
diagnostic feature, and is not covered by the other E-ELT Small
Studies in progress.  Other spectral lines of interest also lie in
this region e.g. the Paschen series of hydrogen.  

In Sections 2.1 to 2.5 we discuss five specific areas of research
that require an instrument such as MOMSI to satisfy the scientific
aims.  Here we have focussed on applications requiring 
spectroscopy.  The science case for imaging\cite{eline} would make
somewhat different demands on the AO system -- this will be more fully
explored in future point-design studies.

\subsection{Stellar kinematics and galactic archaelogy}
Over the past decade new multi-object instrumentation has enabled the
study of so-called `galaxy archaeology'.  This involves mapping the
kinematics of the stellar content of the Milky Way and nearby galaxies
to determine the sub-components of their structure, then combining the
velocity information with abundance analyses of selected stars.  This
fossil record of a galaxy can then be used to investigate the
star-formation history, chemical evolution, evidence for past merger
activity, and the formation of components such as bulges etc.  In
essence, {\it exploring mass assembly at low-redshift}.  A number of
projects are currently exploring the outer components of the Milky Way
(e.g. the RAVE consortium\cite{rave} , the proposed AAOmega ARGUS
programme, the proposed Gemini-WFMOS instrument), with others
focussing their efforts on M31 and its satellites\cite{i04} .

One of the principal limiting factors in ongoing studies remains the
lack of accessible systems -- how typical are the processes that we
see in the Milky Way and the Andromeda group?  With deployable
integral field units (IFUs), MOMSI will be able to examine the
kinematic structure of stellar populations in a wide range of
galaxies.  In particular, probing the less-crowded, outer-regions of a
galaxy can provide some of the most revealing discoveries, as illustrated
by the discovery of relatively faint structures around M31\cite{f02} .
MOMSI will be able to study the kinematics of the most luminous stars
in the outer regions of galaxies in the Virgo Cluster, as well as
obtaining first-order estimates of their chemical compositions.  With
the benefit of a large primary aperture, studies of the halo
populations in dark-matter dominated galaxies at the edge of the Local
Group such as NGC\,3109\cite{m99} will also become feasible.

\subsection{Clusters and starbursts}
M82 is the nearest starburst galaxy, with considerable ongoing star
formation.  Much of this is seen to be occurring in young, very
compact star clusters, usually referred to as super star clusters
(SSCs).  With the aim of understanding significant star-formation in
the early Universe, spectral synthesis techniques have been
applied to SSCs in M82 to understand their mass functions, and the
feedback of ionizing photons into the local intercluster medium.

A `top-heavy' mass function was found for an SSC in M82 \cite{sg01} ,
leading to suggestions of an abnormal initial mass function (IMF) therein.
Such conclusions are still widely debated.  For instance, no evidence
for similar behaviour at low-masses is seen in an SSC in NGC\,1705 at
5~Mpc \cite{v04} .  Resolved (or near-resolved) observations with
MOMSI of such clusters would enable the true stellar population to be
determined, allowing studies of the IMF in each cluster, and the
overall cluster mass function that modulates the IMF.  Given the large
quantity of dust and gas present in these compact regions,
observations in the $K$ band will be crucial.

\begin{figure}[h]
\begin{center}
\includegraphics[height=4cm]{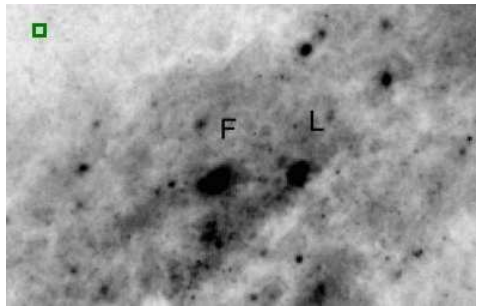}
\end{center}
\caption{Super star clusters in M82 from F814W observations with WFPC2 on the {\it HST}
(from Smith \& Gallagher, 2001).  The on-sky footprint of the MOMSI IFUs
($\sim$0.4x0.4'') is marked with the solid-lined box at the top left of the
image.  Multi-IFU observations in a galaxy such as M82 are perfect
targets for an ELT.}
\label{m82}
\end{figure}

\subsection{Stellar astrophysics in new environments}
One of the main drivers in stellar astrophysics over the past decade
has been to understand the role of metallicity (i.e. abundance of
metallic species) and environment on stellar evolution.  Much of this
has been motivated by a desire to better understand the evolution
and formation of stars in the early Universe.  New large-scale,
high-resolution stellar surveys in the Galaxy and Magellanic Clouds
\cite{flames} have statistically meaningful samples to
rigorously examine the dependence of physical parameters, abundance
enhancements and rotation rates on environment.  Meanwhile,
intermediate-resolution observations of the most visually luminous
stars (which are A-type supergiants) have pushed further out to the
edge of the Local Group, finding galaxies with lower metal abundances, 
e.g. Sextans A\cite{kaufer} .

The high-resolution mode of MOMSI would provide observations of large
samples of the most massive stars in systems such as Sextans A, and beyond.
This would test theoretical model atmospheres and stellar feedback in an 
unexplored metallicity regime -- of interest in the context of the
first stars in the Universe (so-called Population~III stars).   Similar surveys of
red giants and stars on the asymptotic giant branch (AGBs) would also be
of considerable interest to the community.  Theoretical modeling of lines in the 
near-IR has advanced in recent years \cite{tamara} and lack of access
to optical diagnostics would not hamper such efforts.

The direct descendents of massive O-type stars are Wolf-Rayet (WR)
stars.  New studies have discussed the metallicity dependence
of their outflows\cite{vink,ch} , and they have attracted interest as the
likely precursors of some long gamma-ray bursts in low metallicity
systems \cite{eld} .  The discovery that I~Zw~18 is a very metal-poor,
young galaxy\cite{izo} provides an exciting prospect in the era of
ELTs.  Spectroscopic observations of WR stars in systems such as
I~Zw~18 would enable quantitative studies of metal-poor WR stars in
genuinely young, star-forming galaxies.  This would improve our
understanding of these young and violently-evolving stars, and enhance
empirical libraries used to interpret integrated spectra of distant WR
galaxies.

\subsection{Mass assembly at high redshift}
The majority of stellar mass is believed to build-up in galaxies
between $z \sim$1-3.  Significantly massive components are seen within some
galaxies\cite{tec} , and the optical morphology of sub-mm galaxies is
found to be wide-ranging\cite{pope} .  VLT-KMOS will measure
spatially-averaged rotation curves for some of these galaxies but
will be limited by spatial resolution, leading to uncertainties in the
inferred properties.  The long-standing debate
regarding secular (i.e. external events) versus passive evolution of
such galaxies will likely be better understood by the time an E-ELT is
built.  Nevertheless, high spatial resolution observations of these systems
with MOMSI would probe unprecedented scales of distant galaxies -- investigating
both their dynamical structure and chemical compositions.

\subsection{Super-massive blackholes and their hosts}
Massive black holes are ubiquitous components of galaxies, with their
mass (M$_{\rm BH}$) related to the stellar mass of the surrounding bulge/spheroid 
\cite{m98,hr05} (the Magorrian relation).  
M$_{\rm BH}$ and the luminosity-weighted line-of-sight velocity
dispersion ($\sigma$) are also found to be correlated \cite{g00} .
Both the Magorrian and M$_{\rm BH} - \sigma$ correlations represent a
fossil record of past activity and imply that central black hole mass
is determined by, and closely related to, the bulge properties of the
host galaxy.  Both relations are well constrained, with
uncertainties of only 0.3 dex, largely attributable to measurement
errors.  Many plausible models have been put forward to explain the
relations, but it is clear that a physical understanding of the
interaction between a black hole, its accretion disk and the inner
bulge of its host galaxy is required.  With high spatial-resolution
MOMSI will be able to study the dynamics of the inner regions of targeted
host galaxies, at wavelengths relatively impervious to foreground
obscuration.

For a sub-sample of relatively local galaxies\cite{hr05} ,
this is a very compelling, mono-IFU science case -- especially for
disentangling the stellar population around the central black hole in
M31 \cite{b05} .  Studying the cores of multiple galaxies in more
distant galaxy clusters is also a possibility, e.g. in Abell 2443 at
$z\sim$0.1\cite{t01} .

\section{BASELINE REQUIREMENTS}\label{req}
Following the assumed baseline of the current European working groups,
we adopt a 42-m primary aperture.  In Table~\ref{reqtab} we summarize
the specifications of MOMSI.  Each of these is now discussed in turn:

\begin{itemize}

\item{\bf Patrol field:}  The patrol field for MOMSI is 2$\times$2 
arcmin.  This choice is primarily driven by the spatial extent of
galaxies in the Virgo Cluster, with the proposed field encompassing
roughly a quadrant of a major spiral such as M100 (see
Figure~\ref{m100}).  For smaller galaxies in Virgo it would be
possible to simply pick-off sample regions around the outer parts of
the whole galaxy in one exposure.

\begin{figure}[ht]
\begin{center}
\includegraphics[height=8.5cm]{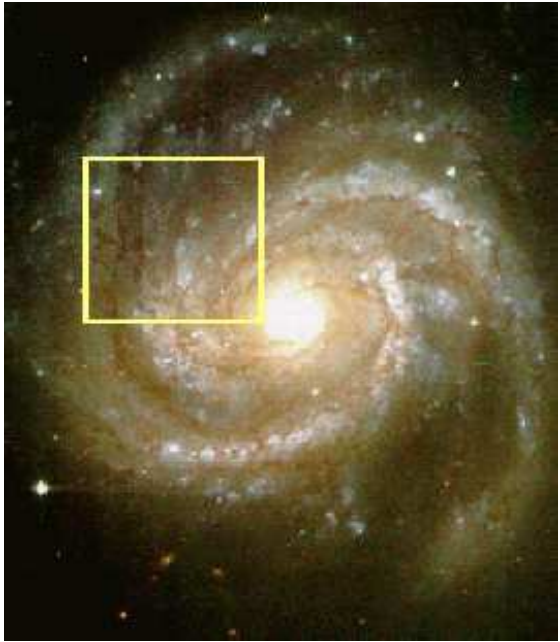}
\end{center}
\caption{M100 in the Virgo Cluster.  This is a combined $BVI$ image from the Isaac Newton
Telescope (INT) in La Palma, with north-east at the top left.  The
solid-lined box indicates the MOMSI patrol field of 2$\times$2 arcmin.}
\label{m100}
\end{figure}

\item{\bf Spectroscopic pick-off FOV:}  The SSCs in 
M82 (Figure~\ref{m82}) are well-matched by a spectroscopic pick-off of
$\sim$0.4-0.5'' squared.  Such scales are also comparable to the
typical half-light radii of $z \sim$3 galaxies (of order 0.2-0.3''
\cite{gsm}).  Allowing for a conservative inter-slice gap, a 2k array
detector can accommodate 40 slices in each IFU, which at 10 mas
sampling gives a FOV of 0.4''$\times$0.4''.  Slightly larger fields could 
be useful but, if working at the highest spatial-resolution, would require larger (or
additional) detectors, image slicers etc.

\item{\bf Imaging pick-off FOV:}  At 5mas sampling, 10''$\times$10'' corresponds to a 
2k array in each pick-off.  To put this in context, at a distance of
16~Mpc (i.e. the Virgo Cluster) this is roughly equivalent to imaging the
stellar content within a 1-degree field in the Large
Magellanic Cloud, which more than adequately samples a wide range of
structures.  

\item{\bf Multiplex:}  
The stellar populations part of the science case would benefit from as
large a multiplex as possible, although studies of high-$z$ galaxies
may not have sufficient source densities in a 2$'$ field to complement
this.  Considering such arguments, in conjunction with physical space
and design constraints (especially those arising from the
AO-system) we adopt an approach with 20 IFU, spectroscopic pick-offs
and 10 imaging pick-offs.

\item{\bf Spatial resolution:} In principle we want to fully exploit 
the spatial information from an E-ELT.  This leads to a goal of 50\%
ensquared energy in 10 mas (in the $H$-band), with critical 2-pixel
sampling for imaging, though this is less pressing in the
spectroscopic mode.

\item{\bf Spectral resolution:} The science case requires two spectroscopic 
modes -- `low-resolution' with $R \sim$~4,000 (to resolve the
sky emission lines in the near-IR) and `high-resolution' with
$R \sim$~20,000.  The requirement for the high-resolution mode comes from the
desire to determine accurate chemical abundances, using
metallic lines that are often relatively weak and
blended.  Past experience with near-IR instruments (e.g. NIRSPEC on
Keck II) has demonstrated the suitability of this resolution for
accurate abundances\cite{orc} ; similarly in the optical with 
VLT-FLAMES\cite{ian} .

\item{\bf Wavelength range:}  $JHK$ required, with $I$ and $Z$ bands as a goal.
At $R \sim$~4,000, full coverage of a passband is required
in one exposure (i.e. $J$, $H$, $K$, or $I+Z$).  In the
high-resolution mode, a 2k array detector will give coverage of $\sim$0.05 to 0.2~$\mu$m in one
exposure, depending on the central wavelength.  This is adequate 
to, for example, determine abundances in cool, red giant stars which have a
large number of metallic lines in the $H$ band \cite{orc} .

\end{itemize}

\begin{table}[h]
\caption{Summary of baseline requirements for MOMSI.}\label{reqtab}
\begin{center}
\begin{tabular}{|p{3cm}|p{2.5cm}|p{2.5cm}|}
\hline
{\bf Requirement} & {\bf Imaging} & {\bf Spectroscopy} \\
\hline
Patrol field & \multicolumn{2}{c|}{2'$\times$2'}  \\
\hline
Pick-off FOV & 10$\times$10 arcsec & 0.4$\times$0.4 arcsec \\
\hline
Multiplex & 10 & 20 \\
\hline
Spatial resolution & \multicolumn{2}{c|}{50\% of energy within 10 mas} \\
\hline
Spatial sampling & \multicolumn{2}{c|}{5 mas} \\
\hline
Spectral resolution & $-$ & 4,000 \& 20,000 \\
\hline
Wavelength range & \multicolumn{2}{c|}{(0.8) 1.0--2.5$\mu$m} \\
\hline
\end{tabular}
\end{center}
\end{table}

\section{MULTI-OBJECT ADAPTIVE OPTICS}\label{ao}

The conceptual design for MOMSI relies on AO-correction from a
multi-object adaptive optics (MOAO) system.  Instead of correcting the
whole focal plane, MOAO corrects the wavefront locally for each
target.  This was the adopted AO system for the proposed VLT-FALCON
concept\cite{fa} .  FALCON uses several off-axis natural guide stars
(NGS) to perform wavefront sensing around each scientific target.
Using atmospheric tomography techniques it is possible to then compute
the best commands to apply to the deformable mirror (DM) in each IFU.
Scaling the concept to an E-ELT requires several laser guide stars
(LGS) in order to solve for the cone effect, as well as NGS to correct
for tip-tilt (and for possibly for windshake effects depending on the
telescope design).

The demands of imaging and spectroscopy on the AO-correction are
somewhat different.  For high-quality imaging, particularly to
disentangle crowded fields, parameters such as Strehl ratio and FWHM
of the AO-corrected core are useful parameters.  For spectroscopy, one
is more concerned with achieving the maximum ensquared energy per
element, rather than paying attention to the actual profile of the
PSF.  One method to improve the ensquared energy (thence the final
signal-to-noise) is to relax the critical spatial sampling requirement such
that the useful aperture (spatial pixel) on the sky is at least 10~mas
($\sim$ the diffraction limit in the $H$ band).  

The coupling factor is defined as the fraction of the energy ensquared
within the useful aperture, of a PSF centered on that aperture
\cite{agh} .  Seeing limited couplings of $\sim$15\% are not sufficient, 
and ensquared energies are required of at least 30\%, and preferably
in the range of 40-50\%.  MOAO remains an unproven technique, but
appears promising for an E-ELT.  One of the potential problems remains
how well the correction of the wavefront can be interpolated using one
set of LGS (say 8-10) across the whole field, i.e. definitely outside
of the isoplanatic patch on the sky.  Depending on the performance of
such methods, a possible scenario could be to have a constellation of
lasers within the isoplanatic patch for one central pick-off, giving a
significant coupling factor at the finest spatial resolution.  An
example (mono-IFU) science case for this could be to study the stellar
population around the central black hole in M31 \cite{b05} .  The
other pick-offs could then rely on LGS across the whole field, effectively
working at coarser spatial resolution (perhaps employing
multi-conjugate adaptive optics (MCAO) to correct
across the whole patrol field, rather than MOAO).

\section{OPTO-MECHANICAL OVERVIEW}
With a complex AO system and a high-resolution spectroscopic mode, 
a gravity-stable instrument platform is required.  In what follows we presuppose
that the MOMSI instrument is located on a vertical Nasmyth platform.

\subsection{Separating laser guide stars, natural guide stars, and science targets}
Two approaches are possible:
\begin{itemize}
\item{A large dichroic is placed at some distance from the Nasmyth focal plane, 
reflecting the science and NGS beams (0.8-2.5 $\mu$m) and transmitting
the LGS that are distributed across the 2$\times$2 arcmin
field-of-view.  The number of LGS and NGS is still to be fixed as it
will depend on complex modelling of the AO-concept (Section~\ref{ao}).
Shack-Hartman wave-front sensors then measure the atmospheric
wavefront errors and these are fed to the DMs in each arm.  This
solution is attractive since it defines the interfaces precisely, but
works in open loop.}
\item{Relatively small dichroics are used in parallel beams, placed after the DMs.  In this 
case they could be used in closed loop.  However, this significantly
complicates the focal plane, and likely requires dedicated LGS/NGS per
pick-off.  We do not consider this option further for now.}
\end{itemize}

\subsection{Field derotation}
At Nasmyth focus, both the field and pupil rotate when the telescope
tracks science targets.  There are two possibilities to solve the
field roation problem.  By a mechanical derotator incorporated to the
interface between the telescope and the MOMSI instrument, or by an
optical derotator covering the whole field-of-view.  We have selected
the first option since the necessary optical derotator would be
enormous (2-3m).  For the pupil rotation the DM could rotate, or it
could be included as a software calibration when measuring the
wavefront errors.

\subsection{Atmospheric dispersion corrector}
Although challenging, correction for atmospheric dispersion on an
E-ELT is not insurmountable\cite{thadc} .  A large ADC covering the
whole 2$\times$2 arcmin field-of-view is included, based on two
rotating wedged prisms made of optical materials (e.g. ZnS/ZnSe) to
compensate for the atmospheric chromatic aberration.  If the ADC
performance is not satisfactory to obtain near-diffraction limited
images, small ADCs could be incorporated in the collimated
beams of each pick-off arm.

\subsection{Foreoptics}
\begin{figure}[h]
\begin{center}
\includegraphics[width=9.25cm]{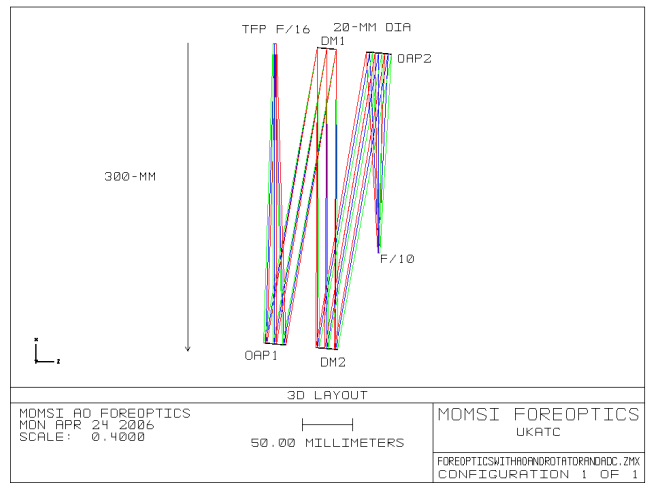}
\end{center}
\caption{Foreoptics in a MOMSI pick-off arm.}
\label{fore}
\end{figure}

The focal plane adopts a similar philosophy to that used in
VLT-KMOS\cite{rs} , i.e. 20 pick-off arms will be used to select
targeted sub-fields and direct the beams to 20 spectrometers.  The 10
imaging pick-off arms will be longer, leading to 10 imagers beyond the
spectrometers.  The foreoptics are shown in Figure~\ref{fore} and
include two DMs to facilitate the wavefront correction from the AO
system.

\subsection{IFU image slicers}
The IFU is designed to slice a 0.4$\times$0.4 arcsec image into 40
slices and reformat them to form a single slit-image at the input to
the spectrograph.  We propose to use a design based on those in
the UKIRT 1-5 micron Imager Spectrometer (UIST)\cite{uist} and the Mid-infrared
Instrument (MIRI)\cite{miri} for the {\it JWST}.

The IFU foreoptics magnify the input image onto a slicing mirror,
consisting of 40 slices.  Each slice is a spherical surface, sharing a
common radius of curvature.  The slicing mirror forms a set of 40 pupil
images that each contain the light from a single slice.  A set of 40
reimaging mirrors are then used to form an image of the slices at the
input focal-plane of the spectrograph, simultaneously reimaging each
of the 40 separate pupils onto a single output pupil at infinity,
forming a telecentric output.

The slicing mirror will consist of 40$\times$20mm long slices.  
Each slice will be reimaged to form a slit image 40 pixels long on the
detector.  To ensure that the spectrum is adequately (i.e. Nyquist)
sampled, each slice image must be a minimum of 2 pixels wide.  The
physical width of the slices also determines the spatial sampling of
the image on the sky.  If the spatial sampling is to be equal in both
dimensions then anamorphic foreoptics must be used to magnify the
image by different factors in the two dimensions. The f/10 input beam
will be converted to f/245 along the slices, giving a plate scale of
10mas = 0.5mm = 1 pixel at detector, and to f/490 across the slices,
giving a plate scale of 10mas = 1mm = 1 slice width.  The reimaging optics after the
slices form demagnified images of the slices, forming a telecentric
output beam which is f/10 along the slices and f/20 across the slices.

Alternatively, simpler foreoptics can be used that magnify both 
dimensions equally, giving different spatial sampling along and across
the slices.  This gives non-square pixels in the reconstructed images, but
has the advantage that the IFU foreoptics will probably consist of a 
single spherical mirror (cf. anamorphic magnification, in which the foreoptics
would include a minimum of two toroidal surfaces).  The main disadvantage
is that the image would be either under- or over-sampled in one dimension.

\subsection{Spectrometers}
\begin{figure}[ht]
\begin{minipage}{8.5cm}
\begin{center}
\includegraphics[width=8.25cm]{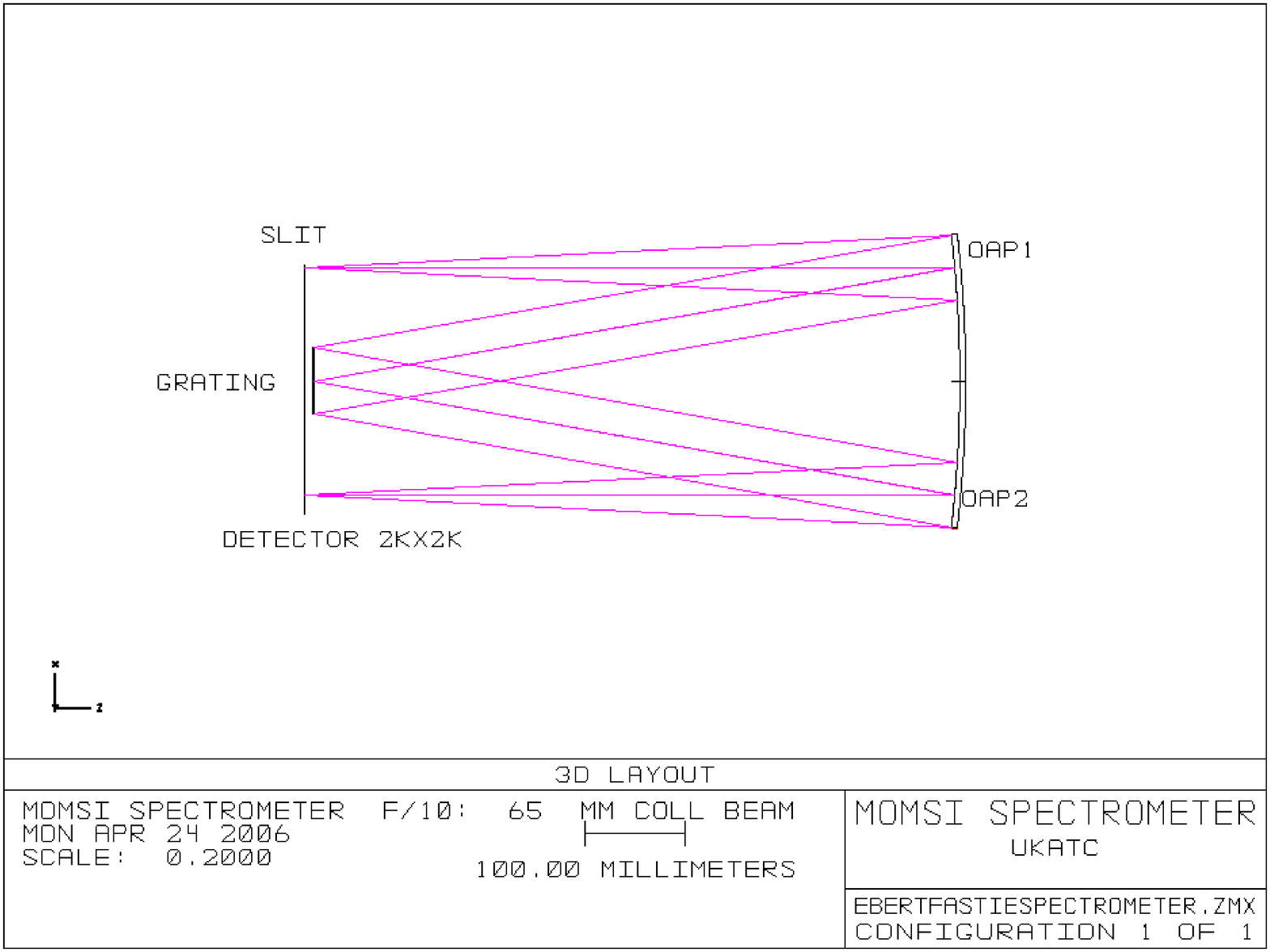}
\caption{Ebert-Fastie spectrometer (XY-plane).}\label{spec_xy}
\end{center}
\end{minipage}
\begin{minipage}{8.5cm}
\begin{center}
\includegraphics[width=8.25cm]{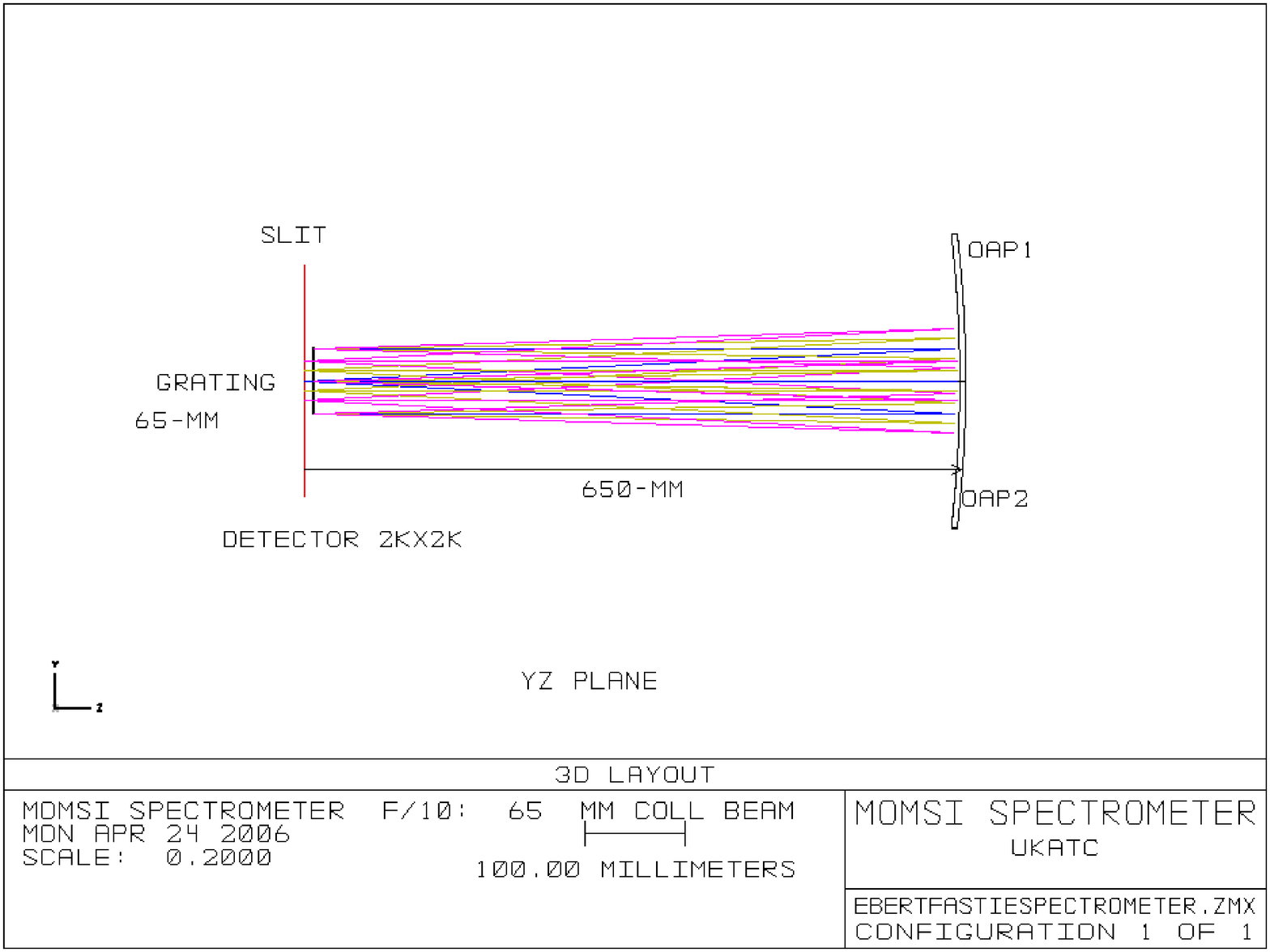}
\caption{Ebert-Fastie spectrometer (YZ-plane).}\label{spec_yz}
\end{center}
\end{minipage}
\end{figure}

The spectrometers are an Ebert-Fastie design (input and output f/10),
operating over a wavelength range of 0.8-2.5~$\mu$m, with a
2k$\times$2k detector array.  Two gratings provide the different
spectral resolutions.  In the high-resolution mode, the
critically-sampled coverage in one exposure will range from 0.05 to
0.2$\mu$m, depending on the central wavelength.  In the low-resolution
mode, either $J$, $H$, $K$, or $I+Z$ is covered in an exposure.  The
optical concept is shown in Figures~\ref{spec_xy} and
\ref{spec_yz}.

\begin{figure}[h]
\begin{center}
\includegraphics[height=4cm]{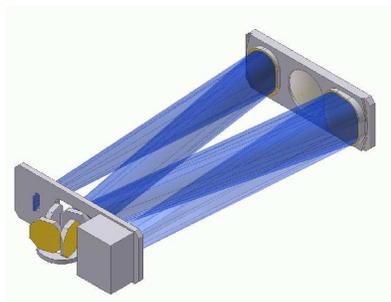}
\caption{Mechanical design of spectrometer unit.}
\label{spec}
\end{center}
\end{figure}

\begin{figure}[ht]
\begin{center}
\includegraphics[height=10cm]{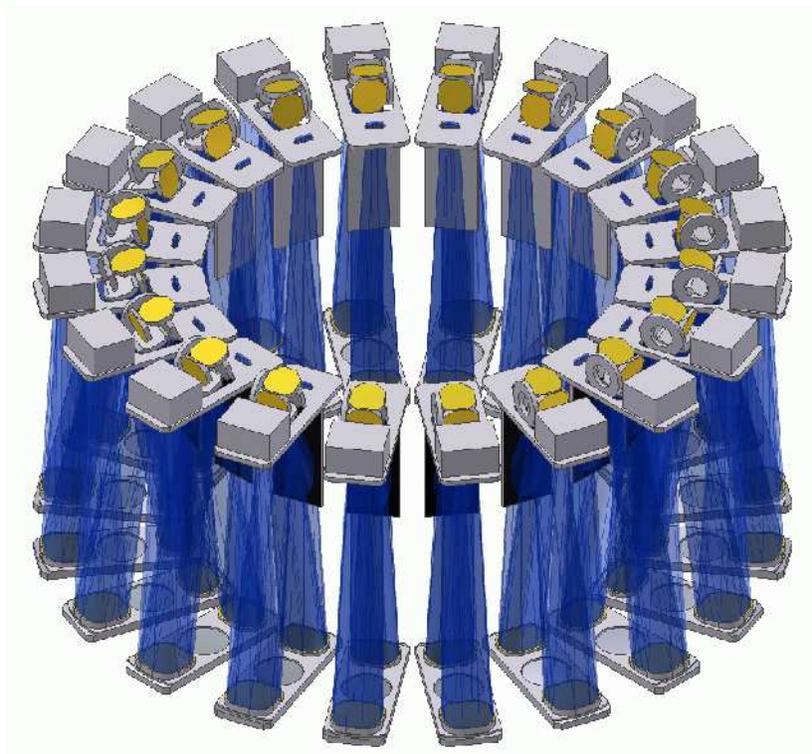}
\caption{Spectrometer layout.}
\label{layout}
\end{center}
\end{figure}

The mechanical concept for the spectrometer units is shown in
Figure~\ref{spec}.  Each spectrometer is 600mm long, 360mm wide and
130mm deep. The collimator and camera mirrors are identical off-axis
parabolas, cut from a single substrate.  The gratings are used in
first-order and are mounted on a turntable to switch between the two
resolutions.  The ring of 20 spectrometers is shown in
Figure~\ref{layout} and has an outer diameter of 1600mm.

\subsection{Imagers}
\begin{figure}[ht]
\begin{center}
\includegraphics[width=8.25cm]{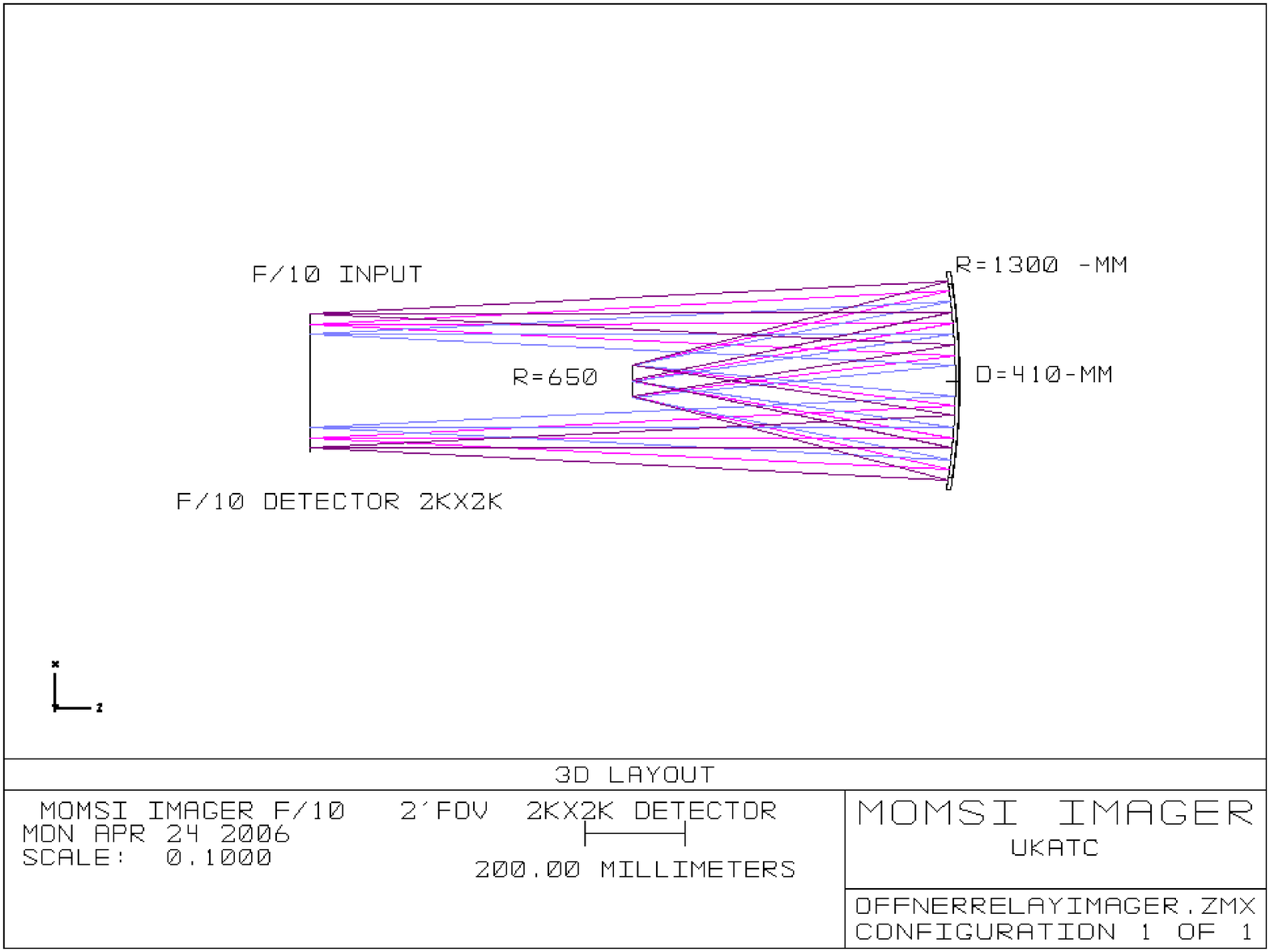}
\caption{Offner relay imager.}
\label{imager}
\end{center}
\end{figure}

The imagers employ an Offner relay design (as shown in
Figure~\ref{imager}), with a 2k$\times$2k detector array.  These will
also operate over a wavelength range of 0.8-2.5~$\mu$m.

\section{SUMMARY}
We have presented an initial design concept for a multiplexed near-IR
instrument for an E-ELT, together with illustrative science cases for
resolved stellar populations and galaxy evolution.  The concept
presented here employs pick-off arms similar to those developed for
VLT-KMOS.  These direct the science sub-fields to DMs for
AO-correction, then onto image slicers for IFU spectroscopy, or to
imaging cameras.

As with other ELT studies, the area in which the most development and
effort is required is with regard to expected AO performance.  As the
programme for an E-ELT advances, further simulations of MOAO are
required -- in terms of the number of LGS/NGS required for a
given AO-correction, and also in conjunction with modelling of potential science
targets e.g. consideration of crowding effects in
distant stellar populations.  These will be incorporated into
detailed point-design studies in the near future.

\acknowledgments      
This activity is supported by the European Community (Framework
Programme 6, ELT Design Study, Contract Number 011863).  We thank the
staff at the Isaac Newton Group in La Palma for use of the M100 image.

\bibliography{spie}   
\bibliographystyle{spiebib}   

\end{document}